\documentclass[aps,preprint]{revtex4}%
\usepackage{amsfonts}
\usepackage{amsmath}
\usepackage{amssymb}
\usepackage{graphicx}%
\providecommand{\U}[1]{\protect\rule{.1in}{.1in}}
\begin{document}
\title{Effects of Sample Specific Variations and Fluctuations of Thermal Occupancy on
Fluctuations of Thermodynamic Quantities}
\author{R. A. Serota}
\affiliation{Department of Physics, University of Cincinnati, Cincinnati, OH 45244-0011, serota@physics.uc.edu}


\begin{abstract}
Standard results for the fluctuations of thermodynamic quantities are derived
under the assumption of sampling identical systems that are in different, not
fully equilibrated states. These results apply to fluctuations with time in a
particular macroscopic body and can be traced to the fluctuations of thermal
occupancy. When many identically prepared - but not identical - systems are
studied, mesoscopic fluctuations due to variations from sample to sample
contribute to the fluctuations of thermodynamic quantities. We study the
combined effect of mesoscopic fluctuations and fluctuations of thermal
occupancy. In particular, we evaluate the total particle number and specific
heat fluctuations in a two-dimensional, non-interacting electron gas in
classically integrable and chaotic circumstances.

\end{abstract}
\date{07-23-2008}
\maketitle

\section{Introduction}

The state of thermal equilibrium of a closed system is achieved when its
entropy is at maximum. In this state, the thermodynamic quantities
characterizing macroscopic bodies that comprise it are at their mean values
(thermodynamic averages).\cite{R1} Thermodynamic fluctuations are the
deviations of these quantities from their mean values and are described by the
Einstein formula for the probability density\cite{LL}:
\[
w\left(  x\right)  =const\exp\left[  S\left(  x\right)  \right]
\]
where $S\left(  x\right)  $ is the entropy and $x=\left\{  x_{1},x_{2}%
,\ldots,x_{n}\right\}  $ are the thermodynamic quantities. Ultimately, these
fluctuations can be related to the fluctuations of thermal occupancy. For
instance, thermal occupancy of quantum gases can be derived starting with a
generally non-equilibrium state, followed by maximization of entropy subject
to constraints (conserved quantities).\cite{LL}

This approach to fluctuations, however, neglects sample-specific variations:
it assumes that sampling occurs in the ensemble of identical systems that are
in different states, slightly away from the equilibrium.\cite{R2} On the other
hand, it has become abundantly clear over the last half-century that
sample-specific variations are an important source of fluctuations, as
witnessed by developments in mesoscopic physics. These must be included when
thermodynamic properties of many identically prepared, yet not identical,
systems are studied. In this work we investigate the combined contribution of
mesoscopic fluctuations and fluctuations of thermal occupancy to fluctuations
of thermodynamic quantities.

We use a simple formalism that gives fluctuations of thermodynamic quantities
as a sum of two terms: the first due to thermal occupancy fluctuations and the
second due to sample-specific fluctuations. The latter is expressed in terms
of the correlation function of the density of energy levels. The results are
illustrated for the fluctuations of the number of particles and of the
specific heat in systems with the fixed chemical potential and volume. As a
model system, we use the degenerate, two-dimensional, non-interacting electron
gas with level statics corresponding to classically chaotic and integrable
motions respectively.

\section{Formalism}

In the most general form, the value of a thermodynamic quantity, whose density
is expressed as a function of energy, can be computed as the integral over the
spectrum
\begin{equation}
G=\int d\varepsilon\rho\left(  \varepsilon\right)  f\left(  \varepsilon
\right)  g\left(  \varepsilon\right)  \label{G}%
\end{equation}
where $\rho\left(  \varepsilon\right)  $ level density and $f\left(
\varepsilon\right)  $ is the thermal occupancy. Under an obvious assumption
that $\rho$ and $f$ are uncorrelated variables, we find the following
expressions for the mean and the variance
\begin{equation}
\overline{G}=\int d\varepsilon\overline{\rho}\left(  \varepsilon\right)
\overline{f}\left(  \varepsilon\right)  g\left(  \varepsilon\right)
\label{Gav}%
\end{equation}%
\begin{align}
\overline{\delta G^{2}}  &  =\overline{\delta G_{\rho}^{2}}+\overline{\delta
G_{f}^{2}}\label{Gvar}\\
\overline{\delta G_{\rho}^{2}}  &  =\int\int d\varepsilon_{1}d\varepsilon
_{2}\overline{\delta\rho\left(  \varepsilon_{1}\right)  \delta\rho\left(
\varepsilon_{2}\right)  }\overline{f}\left(  \varepsilon_{1}\right)
\overline{f}\left(  \varepsilon_{2}\right)  g\left(  \varepsilon_{1}\right)
g\left(  \varepsilon_{2}\right) \label{Gvar-rho}\\
\overline{\delta G_{f}^{2}}  &  =\int\int d\varepsilon_{1}d\varepsilon
_{2}\overline{\rho}\left(  \varepsilon_{1}\right)  \overline{\rho}\left(
\varepsilon_{2}\right)  \overline{\delta f\left(  \varepsilon_{1}\right)
\delta f\left(  \varepsilon_{2}\right)  }g\left(  \varepsilon_{1}\right)
g\left(  \varepsilon_{2}\right)  \label{Gvar-f}%
\end{align}
where the over-bar stands for a mean value.\ Using
\begin{equation}
\delta f\left(  \varepsilon\right)  =\frac{\partial\overline{f}\left(
\varepsilon\right)  }{\partial\mu}\left(  \frac{\partial\overline{N}}%
{\partial\mu}\right)  ^{-1}\delta N+\frac{\partial\overline{f}\left(
\varepsilon\right)  }{\partial T}\delta T \label{f-delta}%
\end{equation}
where $\mu$ is the chemical potential and $T$ is temperature, we find
\begin{equation}
\overline{\delta f\left(  \varepsilon_{1}\right)  \delta f\left(
\varepsilon_{2}\right)  }=\frac{\partial\overline{f}\left(  \varepsilon
_{1}\right)  }{\partial\mu}\frac{\partial\overline{f}\left(  \varepsilon
_{2}\right)  }{\partial\mu}\left(  \frac{\partial\overline{N}}{\partial\mu
}\right)  ^{-2}\overline{\delta N_{f}^{2}}+\frac{\partial\overline{f}\left(
\varepsilon_{1}\right)  }{\partial T}\frac{\partial\overline{f}\left(
\varepsilon_{2}\right)  }{\partial T}\overline{\delta T^{2}} \label{f_corr}%
\end{equation}
Combining eqs. (\ref{Gav}), (\ref{Gvar-f}) and (\ref{f_corr}), we find, as
expected from $\delta G_{f}=\left(  \partial\overline{G}/\partial N\right)
\delta N+\left(  \partial\overline{G}/\partial T\right)  \delta T$,
\begin{align}
\overline{\delta G_{f}^{2}}  &  =\left(  \frac{\partial\overline{G}}%
{\partial\mu}\right)  ^{2}\left(  \frac{\partial\overline{N}}{\partial\mu
}\right)  ^{-2}\overline{\delta N_{f}^{2}}+\left(  \frac{\partial\overline{G}%
}{\partial T}\right)  ^{2}\overline{\delta T^{2}}\label{Gvar2-f}\\
&  =T\left(  \frac{\partial\overline{G}}{\partial\mu}\right)  ^{2}\left(
\frac{\partial\overline{N}}{\partial\mu}\right)  ^{-1}+\left(  \frac
{\partial\overline{G}}{\partial T}\right)  ^{2}\frac{T^{2}}{\overline{C}}%
\end{align}
where $\overline{C}$ is the mean specific heat and we used the well-known
results from the theory of thermodynamic fluctuations\cite{LL}
\begin{equation}
\overline{\delta N_{f}^{2}}=T\frac{\partial\overline{N}}{\partial\mu
}\text{,\ \ \ \ \ }\overline{\delta T^{2}}=\frac{T^{2}}{\overline{C}}
\label{Nvar-f_Tvar-f}%
\end{equation}
to obtain the second equality. Applying this result to the fluctuations of the
specific heat, for instance, we find
\begin{equation}
\overline{\delta C_{f}^{2}}=T\left(  \frac{\partial\overline{C}}{\partial\mu
}\right)  ^{2}\left(  \frac{\partial\overline{N}}{\partial\mu}\right)
^{-1}+\left(  \frac{\partial\overline{C}}{\partial T}\right)  ^{2}\frac{T^{2}%
}{\overline{C}} \label{Cvar2D-f}%
\end{equation}

\section{Fluctuations in a 2D electron gas}

We apply the above results to a non-interacting, degenerate, two-dimensional
electron gas occupying area $A$. In this case
\begin{equation}
\overline{\rho}\left(  \varepsilon\right)  =\frac{\partial\overline{N}%
}{\partial\mu}=\frac{mA}{2\pi\hbar^{2}}=\Delta^{-1} \label{rhoav}%
\end{equation}
is a constant inverse average level spacing and $m$ is the electron mass.
Using standard low-temperature (Sommerfeld) expansion,\cite{LL} we find $\mu
$-$\varepsilon_{F}$ $\cong-T\exp\left(  -\varepsilon_{F}/T\right)  $, where
$\varepsilon_{F}$ is the Fermi energy, and
\begin{equation}
\overline{\delta N_{f}^{2}}=\frac{T}{\Delta}\text{,\ \ \ \ \ }\overline{\delta
C_{f}^{2}}=\overline{C}\simeq\frac{\pi^{2}}{3}\frac{T}{\Delta}
\label{Nvar2D-f_Cvar2D-f}%
\end{equation}
Next, we proceed to evaluate $\overline{\delta N_{p}^{2}}$ and $\overline
{\delta C_{p}^{2}}$ for the classically chaotic and integrable circumstances
respectively and compare those to $\overline{\delta N_{f}^{2}}$ and
$\overline{\delta C_{f}^{2}}$.

To simplify calculation, we assume a unitary ensemble (broken time reversal
symmetry) in the classically chaotic case\cite{BFFMPW}%
\begin{equation}
\overline{\delta\rho\left(  \varepsilon\right)  \delta\rho\left(
\varepsilon+\omega\right)  }=\frac{\delta\left(  \omega\right)  }{\Delta
}-\frac{\sin^{2}\left(  \pi\omega/\Delta\right)  }{\pi\omega^{2}%
}\label{rho_rho-ch}%
\end{equation}
We also use a simplified ansatz in the classically integrable case\cite{WGS}
\begin{equation}
\overline{\delta\rho\left(  \varepsilon\right)  \delta\rho\left(
\varepsilon+\omega\right)  }=\frac{\delta\left(  \omega\right)  }{\Delta
}-\frac{\sin\left(  2\pi\omega/E_{m}\right)  }{\pi\omega\Delta}%
\label{rho_rho-in}%
\end{equation}
Here $E_{m}$ is the the energy scale corresponding to the shortest classical
periodic orbit.\cite{B} In a square well, for instance, $E_{m}=\sqrt
{\pi\varepsilon\Delta}$. The simplified ansatz neglects large oscillations of
the density correlation function for $\varepsilon>>E_{m}$.\cite{WGS} These
should be of little consequence at finite temperatures since harmonics with
incommensurate frequencies will be mixed it to wash out the effect of such
oscillations.\cite{WGS} Still, unlike in the chaotic case, the correlation
function depends on $\varepsilon$ via $E_{m}$. However, due to relevant
$\varepsilon$'s being restricted to the interval of order $T$ around
$\varepsilon_{F}$, we can replace $\varepsilon\rightarrow$ $\varepsilon_{F}$
\begin{equation}
E_{m}\approx\sqrt{\pi\varepsilon_{F}\Delta}\label{Em}%
\end{equation}
With this substitution, it will be shown below that for $T\ll E_{m}$ the
result for integrable systems is due to the delta function term in
(\ref{rho_rho-in}). Conversely, for $T\gtrsim E_{m}$, the full oscillatory
behavior of the level correlation function discussed in Ref.\cite{WGS} should
be of little consequence since their scale is defined by $E_{m}$. This
reasoning is ultimately confirmed by numerical evaluation using the full form
of the correlation function\cite{WGS} (with substitution $\varepsilon
\rightarrow$ $\varepsilon_{F}$), which yields results that are extremely close
to the analytical results for $\overline{\delta N_{p}^{2}}$ obtained below
using (\ref{rho_rho-in}).

To eliminate temperature-independent divergencies, we evaluate
\begin{equation}
\frac{\partial\overline{\delta N_{p}^{2}}}{\partial T}=\int\int d\varepsilon
_{1}d\varepsilon_{2}\overline{\delta\rho\left(  \varepsilon_{1}\right)
\delta\rho\left(  \varepsilon_{2}\right)  }\frac{\partial\left[  \overline
{f}\left(  \varepsilon_{1}\right)  \left(  \overline{f}\left(  \varepsilon
_{2}\right)  -1\right)  \right]  }{\partial T} \label{Nvar2D_dT-p}%
\end{equation}
where we used that $\int d\varepsilon_{2}\overline{\delta\rho\left(
\varepsilon_{1}\right)  \delta\rho\left(  \varepsilon_{2}\right)  }=0$.
Changing variables to $\varepsilon=\left(  \varepsilon_{1}+\varepsilon
_{2}\right)  /2$, $\omega=\varepsilon_{2}-\varepsilon_{1}$ and integrating on
$\varepsilon$, we find
\begin{equation}
\frac{\partial\overline{\delta N_{p}^{2}}}{\partial T}=-\int_{-\infty}%
^{\infty}d\omega\overline{\delta\rho\left(  \varepsilon\right)  \delta
\rho\left(  \varepsilon+\omega\right)  }H\left(  \frac{\omega}{T}\right)
\label{Nvar2D_dT1-p}%
\end{equation}
where
\begin{equation}
H\left(  x\right)  \equiv\left[  \frac{x}{2}\operatorname{csch}\left(
\frac{x}{2}\right)  \right]  ^{2} \label{H}%
\end{equation}
We first substitute (\ref{rho_rho-ch}), (\ref{rho_rho-in}) and (\ref{H}) into
(\ref{Nvar2D_dT-p}) and evaluate the integrals analytically. Further
integration on $T$ gives
\begin{equation}
\left(  \overline{\delta N_{p}^{2}}\right)  _{ch}=\left(  \overline{\delta
N_{p}^{2}}\right)  _{ch,T=0}-\frac{T}{\Delta}+\frac{1}{2\pi^{2}}\log\left(
\frac{\sinh\left(  2\pi^{2}T/\Delta\right)  }{2\pi^{2}T/\Delta}\right)
\label{Nvar2D_ch-p}%
\end{equation}
and
\begin{equation}
\left(  \overline{\delta N_{p}^{2}}\right)  _{in}=\left(  \overline{\delta
N_{p}^{2}}\right)  _{in,T=0}-\frac{T}{\Delta}+\frac{T}{\Delta}\coth\left(
\frac{2\pi^{2}T}{E_{m}}\right)  -\frac{E_{m}}{2\pi^{2}\Delta}
\label{Nvar2D_in-p}%
\end{equation}
for classically chaotic and integrable systems respectively, with $E_{m}$
given by (\ref{Em}) and zero-temperature fluctuations given by
\begin{equation}
\left(  \overline{\delta N_{\rho}^{2}}\right)  _{T=0}=\int_{0}^{\varepsilon
_{F}}\int_{0}^{\varepsilon_{F}}d\varepsilon_{1}d\varepsilon_{2}\overline
{\delta\rho\left(  \varepsilon_{1}\right)  \delta\rho\left(  \varepsilon
_{2}\right)  } \label{Nvar2D_0}%
\end{equation}
The latter is simply the variance of the number of levels over the Fermi sea
and we use the results for the level number fluctuations from Ref.\cite{WGS}
to find
\begin{equation}
\left(  \overline{\delta N_{p}^{2}}\right)  _{ch,T=0}\sim\log\left(
\frac{\varepsilon_{F}}{\Delta}\right)  \label{Nvar2D_0-ch}%
\end{equation}
and
\begin{equation}
\left(  \overline{\delta N_{p}^{2}}\right)  _{in,T=0}\sim\frac{E_{m}}{\Delta
}\sim\sqrt{\frac{\varepsilon_{F}}{\Delta}} \label{Nvar2D_0-in}%
\end{equation}
for classically chaotic and integrable systems respectively.%
%TCIMACRO{\FRAME{ftbpFU}{6.2111in}{3.9522in}{0pt}{\Qcb{Function $F\left(
%x\right)  $ given by eq. (\ref{F})}}{\Qlb{Ffig}}{f.eps}%
%width 6.2111in;  height 3.9522in;  depth 0pt;  original-width 6.1834in;
%original-height 3.9245in;  cropleft "0";  croptop "1";  cropright "1";
%cropbottom "0";  filename 'F.eps';file-properties "XNPEU";}} }%
%BeginExpansion
\begin{figure}
[ptb]
\begin{center}
\includegraphics[
height=3.9522in,
width=6.2111in
]%
{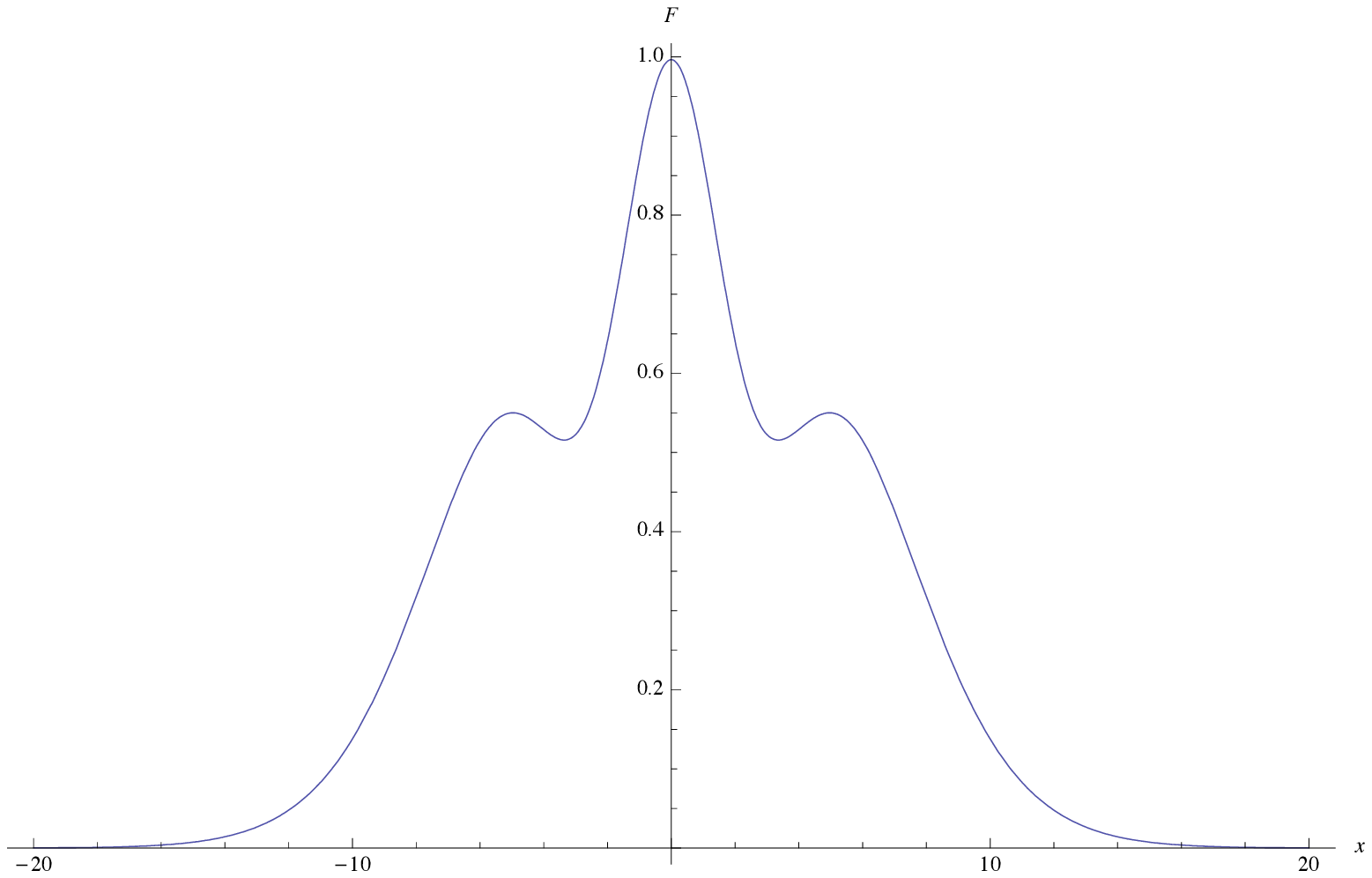}%
\caption{Function $F\left(  x\right)  $ given by eq. (\ref{F})}%
\label{Ffig}%
\end{center}
\end{figure}
%EndExpansion

Combining now with (\ref{Nvar2D-f_Cvar2D-f}), we find for the total particle
number fluctuation $\overline{\delta N^{2}}=\overline{\delta N_{\rho}^{2}%
}+\overline{\delta N_{f}^{2}}$
\begin{align}
\left(  \overline{\delta N^{2}}\right)  _{ch}-\left(  \overline{\delta
N_{p}^{2}}\right)  _{ch,T=0}  &  =\frac{1}{2\pi^{2}}\log\left(  \frac
{\sinh\left(  2\pi^{2}T/\Delta\right)  }{2\pi^{2}T/\Delta}\right)
\label{Nvar2D_ch_t}\\
&  \approx%
\begin{array}
[c]{c}%
\pi^{2}T^{2}/3\Delta^{2}\text{,}\\
T/\Delta-\log\left(  4\pi^{2}T/\Delta\right)  /2\pi^{2}\text{,}%
\end{array}%
\begin{array}
[c]{c}%
T\ll\Delta/2\pi^{2}\\
T\gtrsim\Delta/2\pi^{2}%
\end{array}
\label{Nvar2D_ch_lim}%
\end{align}
and
\begin{align}
\left(  \overline{\delta N^{2}}\right)  _{in}-\left(  \overline{\delta
N_{p}^{2}}\right)  _{in,T=0}  &  =\frac{T}{\Delta}\coth\left(  \frac{2\pi
^{2}T}{E_{m}}\right)  -\frac{E_{m}}{2\pi^{2}\Delta}\label{Nvar2D_in_t}\\
&  \approx%
\begin{array}
[c]{c}%
2\pi^{2}T^{2}/3E_{m}\Delta\text{,}\\
T/\Delta-E_{m}/2\pi^{2}\Delta\text{,}%
\end{array}%
\begin{array}
[c]{c}%
T\ll E_{m}/2\pi^{2}\\
T\gtrsim E_{m}/2\pi^{2}%
\end{array}
\label{Nvar2D_in_lim}%
\end{align}
for classically chaotic and integrable systems respectively. Notice, that for
temperatures below the energy scale where level rigidity sets in\cite{BFFMPW}%
,\cite{WGS} - $\Delta$ for classically chaotic and $E_{m}$ for classically
integrable systems - the main effect of temperature is to \textit{reduce}
sample specific fluctuations of the number of particles relative to the
zero-temperature limit. Furthermore, the temperature-dependent part of such
fluctuations largely cancels the fluctuations (\ref{Nvar2D-f_Cvar2D-f}) due to
thermal occupancy. Incidentally, since in this limit the temperature-dependent
sample-specific fluctuations are dominated by the $\delta$-function term in
(\ref{rho_rho-ch}) and (\ref{rho_rho-in}), the temperature-dependent particle
number fluctuation can be found directly from
\begin{equation}
\frac{\partial\overline{\delta N_{p}^{2}}}{\partial T}=\frac{\partial
}{\partial T}\int_{0}^{\infty}\int_{0}^{\infty}d\varepsilon_{1}d\varepsilon
_{2}\overline{\delta\rho\left(  \varepsilon_{1}\right)  \delta\rho\left(
\varepsilon_{2}\right)  }\overline{f}\left(  \varepsilon_{1}\right)
\overline{f}\left(  \varepsilon_{2}\right)  =\frac{1}{\Delta}\frac{\partial
}{\partial T}\int_{0}^{\infty}d\varepsilon\overline{f}^{2}\left(
\varepsilon\right)  \simeq-\frac{1}{\Delta} \label{Nvar2D_dT2-p}%
\end{equation}
confirming the results in (\ref{Nvar2D_ch-p}) and (\ref{Nvar2D_in-p}).
Conversely, for temperatures above the level rigidity scale, fluctuations due
to thermal occupancy dominate.%
%TCIMACRO{\FRAME{ftbpFU}{6.2993in}{4.0015in}{0pt}{\Qcb{$\left(  \delta
%C_{p}^{2}\right)  _{ch}$ vs. $T/\Delta$, with limiting cases given by eq.
%(\ref{Cvar2D_ch_lim-p}) shown as thin straight lines.}}{\Qlb{Cch}}%
%{cch.eps}{\special{ language "Scientific Word";  type "GRAPHIC";
%width 6.2993in;  height 4.0015in;  depth 0pt;  original-width 6.2379in;
%original-height 3.9522in;  cropleft "0";  croptop "1";  cropright "1";
%cropbottom "0";  filename 'Cch.eps';file-properties "XNPEU";}} }%
%BeginExpansion
\begin{figure}
[ptb]
\begin{center}
\includegraphics[
height=4.0015in,
width=6.2993in
]%
{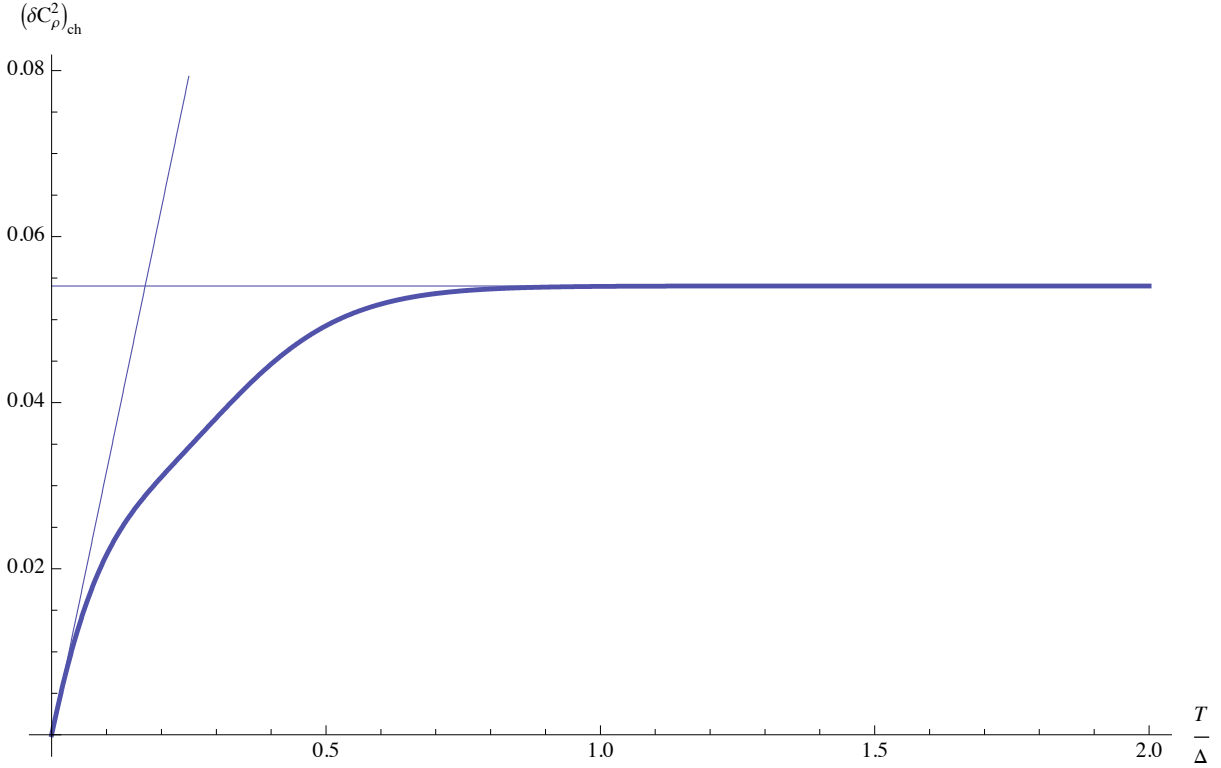}%
\caption{$\left(  \delta C_{p}^{2}\right)  _{ch}$ vs. $T/\Delta$, with
limiting cases given by eq. (\ref{Cvar2D_ch_lim-p}) shown as thin straight
lines.}%
\label{Cch}%
\end{center}
\end{figure}
%EndExpansion
%TCIMACRO{\FRAME{ftbpFU}{6.2439in}{3.9461in}{0pt}{\Qcb{$\left(  \delta
%C_{p}^{2}\right)  _{in}$ vs. $T/E_{m}$, with limiting behavior
%(\ref{Cvar2D_in_lim0-p}) shown as thin straight line. }}{\Qlb{Cin}}%
%{cin.eps}{\special{ language "Scientific Word";  type "GRAPHIC";
%width 6.2439in;  height 3.9461in;  depth 0pt;  original-width 6.1834in;
%original-height 3.8977in;  cropleft "0";  croptop "1";  cropright "1";
%cropbottom "0";  filename 'Cin.eps';file-properties "XNPEU";}} }%
%BeginExpansion
\begin{figure}
[ptb]
\begin{center}
\includegraphics[
height=3.9461in,
width=6.2439in
]%
{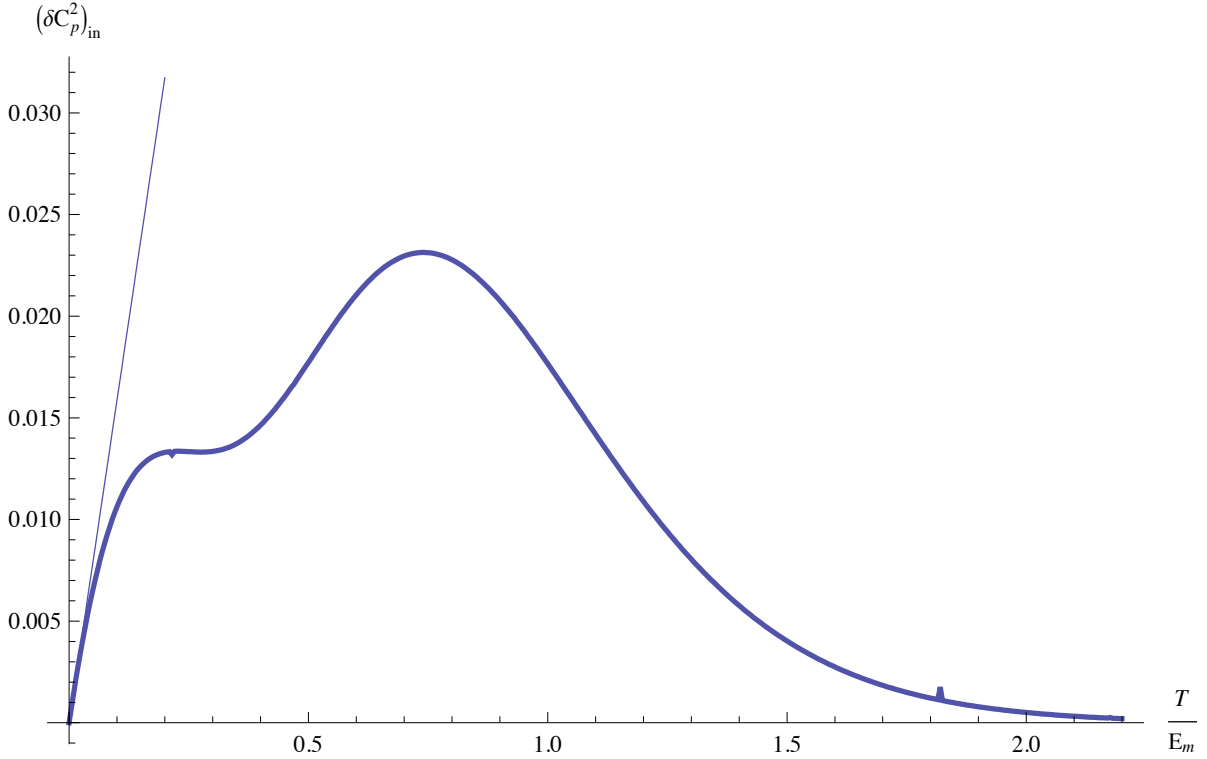}%
\caption{$\left(  \delta C_{p}^{2}\right)  _{in}$ vs. $T/E_{m}$, with limiting
behavior (\ref{Cvar2D_in_lim0-p}) shown as thin straight line. }%
\label{Cin}%
\end{center}
\end{figure}
%EndExpansion

Fluctuations of the specific heat (heat capacity) can be found from\cite{SS}
\begin{equation}
\overline{\delta C_{\rho}^{2}}=\int\int d\varepsilon_{1}d\varepsilon
_{2}\overline{\delta\rho\left(  \varepsilon_{1}\right)  \delta\rho\left(
\varepsilon_{2}\right)  }\frac{\partial\overline{f}\left(  \varepsilon
_{1}\right)  }{\partial T}\frac{\partial\overline{f}\left(  \varepsilon
_{2}\right)  }{\partial T}=T\int\int d\omega\overline{\delta\rho\left(
\varepsilon\right)  \delta\rho\left(  \varepsilon+\omega\right)  }F\left(
\frac{\omega}{T}\right)  \label{Cvar2D-p}%
\end{equation}
where
\begin{equation}
F\left(  x\right)  \equiv\frac{1}{120}\left[  x\left(  14\pi^{4}+x^{4}\right)
\coth\left(  \frac{x}{2}\right)  -4\left(  7\pi^{4}+5\pi^{2}x^{2}\right)
\right]  \operatorname{csch}^{2}\left(  \frac{x}{2}\right)  \label{F}%
\end{equation}
and is plotted in Fig. (\ref{Ffig}). As a result we find the limiting behavior
of the fluctuations
\begin{equation}
\left(  \delta C_{p}^{2}\right)  _{ch}\approx%
%TCIMACRO{\QATOP{aT/\Delta\text{,}}{b\left(  2\pi^{2}\right)  ^{-1}\text{,}}}%
%BeginExpansion
\genfrac{}{}{0pt}{}{aT/\Delta\text{,}}{b\left(  2\pi^{2}\right)  ^{-1}%
\text{,}}%
%EndExpansion%
%TCIMACRO{\QATOP{T\ll\Delta}{T\gtrsim\Delta} }%
%BeginExpansion
\genfrac{}{}{0pt}{}{T\ll\Delta}{T\gtrsim\Delta}
%EndExpansion
\label{Cvar2D_ch_lim-p}%
\end{equation}
where the constants $a$ and $b$ are given by
\begin{subequations}
\begin{equation}
a=F\left(  0\right)  \approx0.997\text{,\ \ \ \ \ }b=\int_{-\infty}^{\infty
}dx\frac{\left[  F\left(  0\right)  -F\left(  x\right)  \right]  }{x^{2}%
}\approx1.067 \label{a_b}%
\end{equation}
and%
\end{subequations}
\begin{equation}
\left(  \delta C_{p}^{2}\right)  _{in}\approx aT/\Delta\text{, \ \ \ \ }T\ll
E_{m} \label{Cvar2D_in_lim0-p}%
\end{equation}
for classically chaotic and integrable systems respectively. Clearly, the
specific heat due to sample-to-sample variations in integrable systems has,
over the large range of temperatures, the same functional dependence as that
due to the fluctuations of thermal occupancy.

We show the plots of $\left(  \delta C_{p}^{2}\right)  _{ch}$ and $\left(
\delta C_{p}^{2}\right)  _{in}$ vs. $T$ in Figs. (\ref{Cch}) and (\ref{Cin})
respectively. Some of the curvature in these plots can be traced to that in
Fig. (\ref{Ffig}). \ Small bumps and dips on the line in Fig. (\ref{Cin}) is
due to poor numerical convergence of rapidly oscillating integrals. For
$T\gtrsim E_{m}$, $\left(  \delta C_{p}^{2}\right)  _{in}$ shows faster than
exponential decay, as expected from integration of an exponentially decaying
function (\ref{F}) with a function rapidly oscillating around zero. In
contrast with chaotic systems, in integrable systems the decay of fluctuations
of the specific heat at high temperature is consistent with the greater level
rigidity at larger energy scales, as discussed in Ref.\cite{WGS}.

\section{Conclusions}

We showed that the fluctuations of thermodynamic quantities can be separated
into those that are due to sample-to-sample (mesoscopic) fluctuations and
those due to thermal occupancy fluctuations. For a 2D non-interacting
degenerate electron gas, both effects were included in evaluation of the
particle number and specific heat fluctuations for classically chaotic and
integrable cases. In the integrable case, the specific heat fluctuations due
to either effect are comparable over a wide range of temperatures. For the
particle number fluctuation, the temperature-dependent part of mesoscopic
fluctuations is negative and, for temperatures below the energy rigidity
scale, largely cancels the fluctuation due to thermal occupancy fluctuations.
For higher temperatures, the latter dominates the temperature-dependent part
of the fluctuation.

\section{Acknowledgments}

I wish to thank Mike Ma and Bernie Goodman for helpful discussions and Richard
Gass for help with \textit{Mathematica}.


\begin{thebibliography}{9}
\bibitem {R2}In a stationary description; as a function of time, fluctuations
in a given system are understood as due to the presence of the heat bath and
working bodies.
\end{thebibliography}
\end{document}